\documentclass[11pt,twoside]{article}
\usepackage{asp2010}
\usepackage{graphicx}
\usepackage{natbib}

\resetcounters

\begin{document}

\title{Galactic binaries with eLISA}
\author{Gijs~Nelemans,$^{1,2,3}$
\affil{$^1$Department of Astrophysics/IMAPP, Radboud  University Nijmegen, P.O. Box 9010, 6500 GL, The Netherlands}
\affil{$^2$Institute for Astronomy, KU Leuven, Celestijnenlaan 200D, 3001 Leuven, 
Belgium}
\affil{$^3$Nikhef, Science Park 105, 1098 XG Amsterdam, The Netherlands}}

\begin{abstract}
I review what eLISA will see from Galactic binaries -- double stars
with orbital periods less than a few hours and white dwarf (or neutron
star/black hole) components. I discuss the currently known binaries
that are guaranteed (or verification) sources and explain why the
expected total number of eLISA Galactic binaries is several thousand,
even though there are large uncertainties in our knowledge of this
population, in particular that of the interacting AM CVn systems. I
very briefly sketch the astrophysical questions that can be addressed
once these thousands of systems are detected. I close with a short
outline of the electro-magnetic facilities that will come on line
before eLISA will fly and the importance of developing analysis plans
using both electro-magnetic and gravitational wave data.

\end{abstract}

\section{Introduction: Galactic binaries}

The most numerous expected astrophysical sources for eLISA are
Galactic binaries with orbital periods below $\sim$1 hour, in
particular double white dwarf binaries. This was already realised
early on \citep[e.g.][]{eis87,hbw90}, even though hardly any of such
binaries were known at the time. In order to fit in the orbits of such
ultra-compact binaries, the components need to be compact stars: white
dwarfs, neutron stars or black holes. 

By now, several classes of such binaries are know: double white
dwarfs, white dwarf -- neutron star binaries and double neutron star
binaries. In addition, two classes of \emph{interacting} binaries
(rather than the above \emph{detached} binaries) are know in which
either a white dwarf or a neutron star accretes (hydrogen deficient)
material from a white dwarf (like) companion. These are called AM CVn
stars and ultra-compact X-ray binaries \citep[see][for
reviews]{2010PASP..122.1133S,nj06}.

Based on these known systems and the general knowledge of binary
evolution, the expected signals for eLISA have been derived, both from
the individual known systems (when bright enough referred to as
\emph{verification sources} \citealt{sv06}) as well as the total
Galactic population. These will be discussed below, as well as an
outlook into what facilities will likely become available to
astronomers before an eLISA like mission will fly.

\section{Verification sources}

\begin{figure}
\includegraphics[height=0.9\textwidth,angle=-90]{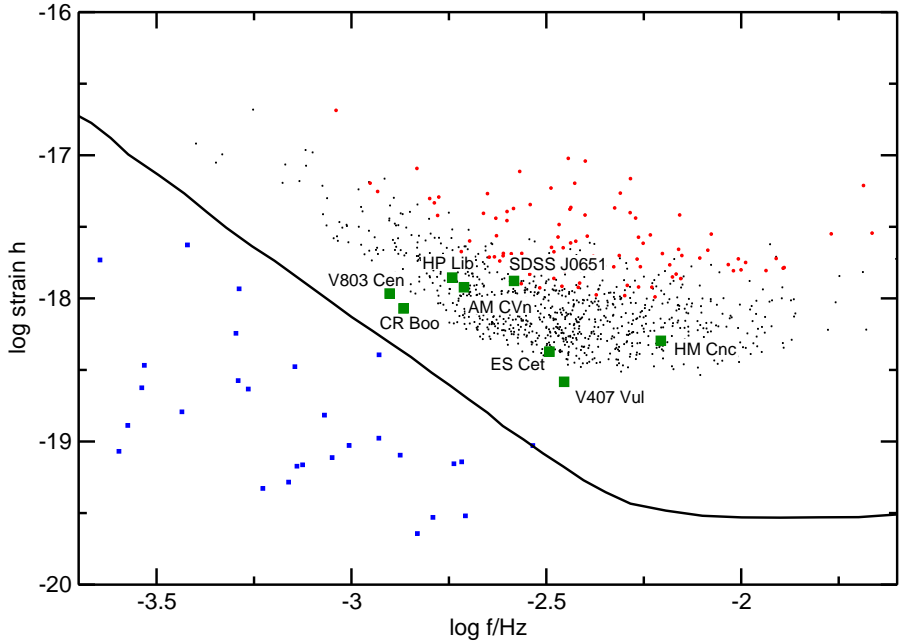}
\label{fig:hf}
\caption[]{Sensitivity plot for eLISA showing the verification
binaries (squares) as well as the thousand brightest expected systems
from a population synthesis model of the total Galactic population of
double white dwarfs. From \citet{2012arXiv1201.3621A}}
\end{figure}

Figure~\ref{fig:hf} shows the expected strain amplitude of the
verification binaries compared to the eLISA sensitivity curve. There
are 8 known binaries that are expected to be detected with S/N above
7: HM Cnc, V407 Vul, ES Cet, SDSS J0651+2844, AM CVn, HP Lib, V803 Cen
and CR Boo.

Compared to earlier reviews
\citep[e.g.][]{2011CQGra..28i4019M,2011ASPC..447..317V,2009CQGra..26i4030N},
there are two important developments. The first is the detection of
the only detached verification binary (yet), SDSS J0651+2844
\citep[][see also Kilic, this volume]{2011ApJ...737L..23B}, for which
recently the orbital decay was measured which is consistent with the
expected rate due to gravitational wave losses
\citep{2012arXiv1208.5051H}.

\begin{figure}
\centerline{\includegraphics[width=0.8\textwidth]{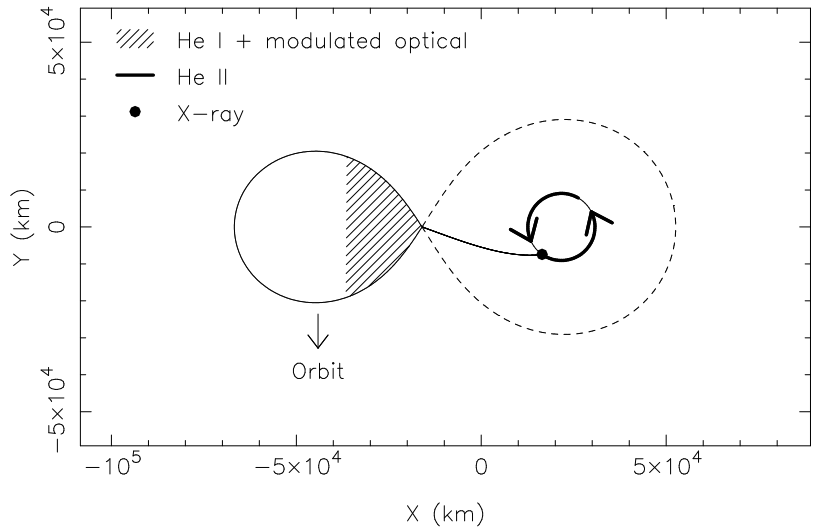}}
\label{fig:HMCnc}
\caption[]{Inferred geometry from the spectral observations that show
counter rotating He I and He II lines. The He II comes from the
material that has hit the accreting white dwarf. The associated X-ray
emission heats the side of the companion facing the accretor, causing
He I emission. The figure is to scale with the axis in km. From
\citet{2010ApJ...711L.138R}}
\end{figure}

The second is the detection of features in the optical spectrum of HM
Cnc that show large velocity variation on the photometric period of
5.4 min \citep{2010ApJ...711L.138R}. This finally decided the debate
on whether this very short photometric period indeed was an orbital
period and confirmed HM Cnc as the brightest verification
binary. Fig.~\ref{fig:HMCnc} shows the inferred geometry (in km!) from
\citet{2010ApJ...711L.138R}.

\section{Expected populations}

\begin{figure}
\includegraphics[width=\textwidth]{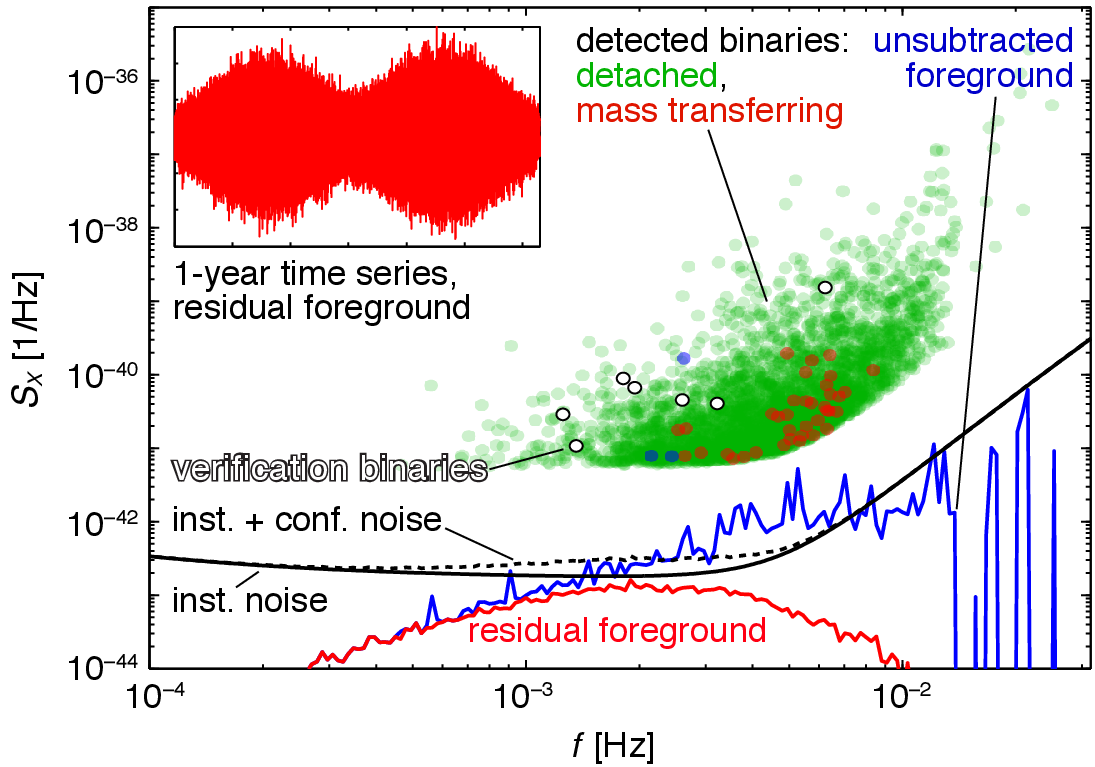}
\label{fig:population}
\caption[]{Summary of expected Galactic binaries for eLISA. The lines
  show the instrument noise, the unsubtracted and subtracted
  foreground and the sum of the instrument plus subtracted foreground
  noise. The points show the expected individual detectable binaries,
  including the versification binaries. From
  \citet{2012CQGra..29l4016A}, based on \citet{2012arXiv1201.4613N}.}
\end{figure}

Based on population synthesis calculations that have (as far as
possible) been calibrated on observed populations, there are some
hundred million double white dwarfs expected in the Galaxy
\citep[e.g.][]{nyp01,2010ApJ...717.1006R,2010ApJ...719.1546L,2010A&A...521A..85Y}. For the original LISA mission, more than
ten thousand systems would be individually detected \citep{trc05,2007PhRvD..75d3008C,2011PhRvD..84f3009L,2011PhRvD..83h3006B}. For the the
eLISA mission the number would be several thousand. In
Fig.~\ref{fig:population}, the expected signal of such a theoretical
population of Galactic binaries is shown.

The main problem for this calculation is that it is known that the
number of AM CVn systems in the Galaxy is significantly smaller than
predicted by these models \citep[][Carter et al. in prep.]{rng07}, at
least at the longer orbital periods where they can be detected in a
homogeneous way. It is unclear what the reason is for this discrepancy
and even simply the whole population is much smaller, e.g. due to
reduced stability at the onset of mass transfer
\citep[see][]{npv+00,mns02} or whether there is something special
happening at longer periods \citep[see][for a more detailed
discussion]{rng07,2012arXiv1201.4613N}. This makes predictions for the
number of AM CVn stars difficult at present. For the detached systems,
the prediction of several thousand individual detections, most with
periods between 5 and 10 minutes is more robust. The number of double
neutron star and/or black hole binaries is much smaller, only several
tens \citep{nyp01,2010ApJ...725..816B}, but for periods below $\sim$30
min eLISA will detect the \emph{whole Galactic population}.

\section{Astrophysical relevance}

\begin{figure}
\includegraphics[height=0.9\textwidth,angle=-90]{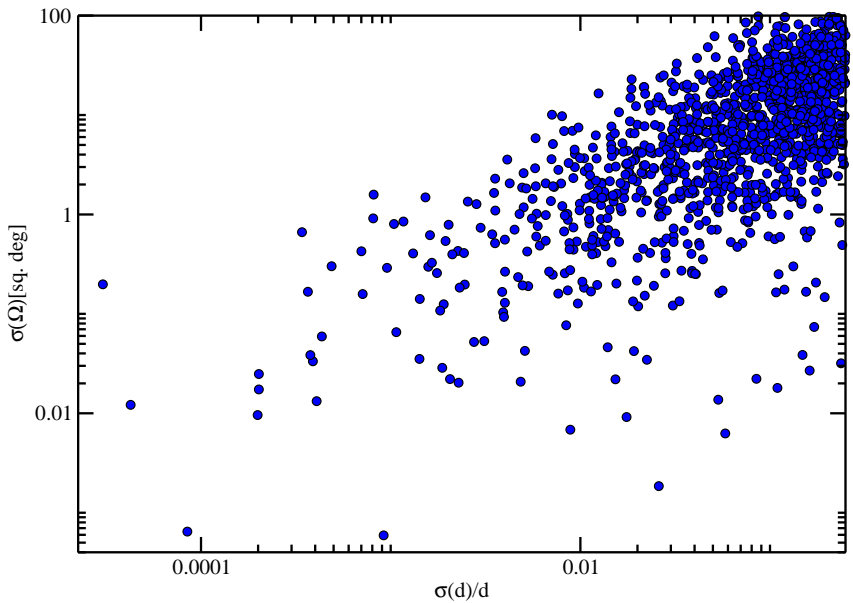}
\label{fig:errors}
\caption[]{Error in distance and sky position for the double white dwarfs individually
  detectable by eLISA. }
\end{figure}

So what will we be able to learn from those thousands of detections
when LISA will fly? Briefly there are the following topics
\begin{description}
\item[Common envelope] All (ultra)-compact binaries in which a white
  dwarf, neutron star or black hole accretes from a low-mass companion
  share one step in their previous evolution. A step in which in some
  way a lot of mass and even more angular momentum is lost from the
  system. The simple idea \citep[][and apparently Ostriker on that
  same conference]{pac76} was that in some way a companion to a giant
  star enters its envelope and then, due to friction, spirals in while
  expelling the envelope. The results is a close binary consisting of
  the core of the giant with the companion. However, it is very
  uncertain how exactly this process happens and indeed in which
  situations it takes place. Double white dwarf binaries can be used
  to constrain the common-envelope phase
  \citep{nvy+00,2012ApJ...744...12W} even with only a hand full of
  systems. The large number of very short period binaries detected by
  eLISA will give strong constraints. For more massive stars, either
  the detection of the double neutron star and/or black hole binaries
  or extrapolation of the results from the white dwarfs might also
  constrain the common-envelope phase.
\item[Type Ia supernova] are used to measure the accelerated expansion
  of the universe, yet we do not know exactly what systems lead to the
  explosion of the white dwarf. One of the proposed progenitor
  scenarios is the merger of two white dwarfs, exactly the population
  that eLISA will map in detail \citep[e.g.][]{2011arXiv1109.4978S}
\item[H deficient accretion (and explosion)] Even if a merger is
  avoided when the two white dwarfs come into contact, or if the
  merger does not lead to a type Ia supernovae, the systems are
  interesting physics laboratories. The (rapid) accretion processes
  that will take place are particularly interesting, because in the
  systems there is no, or hardly any hydrogen. This potentially leads
  to all kinds of interesting phenomena, such as surface explosions,
  double detonations, helium novae and explosive shell ignitions,
  known as .Ia supernovae
  \citep{bsw+07,2009ApJ...706..738W,2010ApJ...709L..64G,2010Natur.463...61P,2010A&A...514A..53F,2011ApJ...737...89D}.
\item[Binary interactions] Even before the double white dwarfs get
  into contact there are potentially other binary interactions. Tidal
  forces may heat the white dwarfs
  \citep[e.g.][]{itf98,2012ApJ...756L..17F,2012MNRAS.421..426F,2012ApJ...745..137V},
  signs of which may be seen in the new 12 minute binary
  \citep{2011ApJ...737L..23B}. This in principle gives information
  about the internal structure of the white dwarfs. Also magnetic
  interactions may influence the orbital evolution.
\item[Galactic structure] Finally, for a subset of the individually
  detectable systems the distance is known (with accuracies randing
  from 10 to less than 1 per cent. Together with reasonable sky
  position errors (Fig.~\ref{fig:errors}) this would allow to build a
  fairly accurate 3D map of where the binaries are in the Galaxy. That
  would give a completely new and different way to look at the
  structure of the Galaxy.
\end{description}

\section{What will happen in the next 10-15 years?}

Now that ESA has decided not to select eLISA for its L1 launch, we
should reconsider the expected progress in the astrophysical questions
in the next 10-15 years. This is difficult, so I will here just
briefly mention the facilities that will come on line and some thoughts
on their influence.

\begin{description}
\item[ALMA] is a large US-European sub-mm array in Chile, that is
  currently performing its first observations. It should be completed
  by 2013 and will be able to make images with superb
  resolution. \texttt{http://www.almaobservatory.org/}
\item[JWST] is the successor of HST, with a 6.5m telescope and
  observing in the infrared. After long delays, the launch date is now
  expected to be 2018. \\\texttt{http://www.jwst.nasa.gov/}
\item[ELTs] are the generic term for the next generation of optical
  telescopes that should significantly increase diameter compared to
  current facilities such as Keck, the VLT, Subaru and Gemini. The
  European E-ELT, with a diameter of 38 meters is planned to become
  operational around
  2022. \\\texttt{http://www.eso.org/public/teles-instr/e-elt.html} The
  US TMT or GMT possibly around 2020 \texttt{http://www.tmt.org/}
  \texttt{http://www.gmto.org/}. These ELTs could be quite
  interesting for Galactic binaries and eLISA because the optical/IR
  signatures of the vast majority of detectable sources will be well
  below capabilities of the current telescope. However, as shown in
  fig.~3 of \citet{2009CQGra..26i4030N}, in the K-bad the majority of
  the systems should have magnitude brighter than about 29, the
  limiting magnitude for an instrument such as MICADO on the E-ELT. 
\item[SKA] the Square Kilometer Array is the next generation radio
  telescope array, that should start construction around 2016. Precursor
  projects such as LOFAR, MeerKAT and ASKAP are currently becoming operational.\\
\texttt{http://www.skatelescope.org/}
\end{description}

\subsection{Transients searches}

The other major development in the next decade will be the
establishment of significant programs to look for transient phenomena
in the sky in the optical/IR regime. Currently projects such as the
Palomar Transient Factory (PTF) and Pan-STARRS are operational and
discovering many strange and new transients. But also know classes of
objects will be discovered at unprecedented rates. PFT, for example,
has already discovered several new AM CVn systems. On the US side,
there is the high-priority plan to build LSST, the ultimate transient
machine by 2022.

\subsection{GAIA}

In 2014, ESA will launch the GAIA satellite which will scan the sky
continuously to determine accurate positions of all stars brighter
than about magnitude 20 and thus determine their parallax and proper
motion \citep[see][for an overview of GAIA]{2008IAUS..248..217L}. GAIA
will be an important source of complementary observations to eLISA. In
particular, the fact that GAIA typically visits each position in the
sky about 80 times, makes it possible to detect eclipsing double white
dwarfs. Because the probability of eclipses increases towards shorter
periods, these are expected to be typically rather short period (less
than a few hours). A preliminary estimate (Marsh \& Nelemans, in prep)
suggest GAIA may detect some 200 of such systems.

\section{Electro-magnetic counterparts}

\begin{figure}
\includegraphics[width=0.99\textwidth]{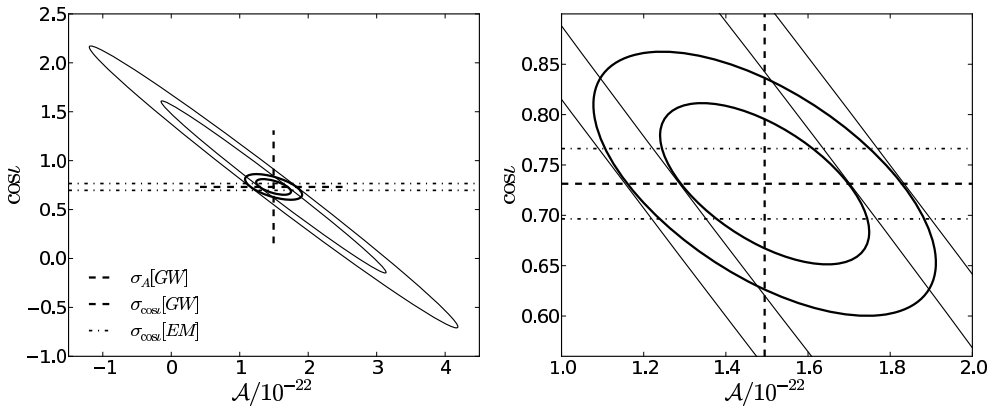}
\label{fig:AMCVn}
\caption[]{Error in amplitude and inclination for AM CVn and the
  improvement in the error on amplitude if the known inclination is
  taken into account. From \citet{2012A&A...544A.153S} }
\end{figure}

Finally, I briefly discuss some first efforts that focus on the
benefit of joint electro-magnetic and gravitational wave data on
individual binaries. For eLISA the number of individually detectable
binaries that can in principle be found by modest size telescopes
(with wide fields of view) is estimated to be between several tens and
several hundreds \citep{2012arXiv1207.4848L}.

\citet{2012A&A...544A.153S} investigated the use of prior
electro-magnetic data for improving parameters that are estimated from
the gravitational wave data. The strongest effect was found to be the
large improvements in determination of the gravitational wave
amplitude (i.e. the combination of masses and distance of the
sources), if the inclination of the source was known from
electro-magnetic observations, like in the verification binary AM CVn
itself (see Fig.~\ref{fig:AMCVn}). The reason is that for the
gravitational wave data there is a very strong correlation between
amplitude and inclination (lower amplitude signals at favourable
[i.e. face on] inclination cannot be distinguished from intrinsically
stronger signals viewed at less optimal inclinations). Interestingly,
this correlation disappears for edge on systems, yielding no
additional gain in determination of the amplitude if the inclination is
known.  This means that, contrary to what one may think, optical data
on eclipsing binaries, where there is the best chance of determining
the inclination from electro-magnetic data, are not so useful.  The
other obvious parameters that are easy to determine from
electro-magnetic data (once the source is found!) is the sky
position. Investigations on how useful this information is are
underway (Shah et al. in preparation). After these first steps a real
integrated approach to joint electro-magnetic and gravitational wave
data analysis should be developed \citep[see also][]{2012arXiv1207.4848L}

\section{Conclusions}

Galactic binaries are also for eLISA a significant source class and
the numbers of individually detectable systems still range in the
thousands. There are currently 8 objects know that would serve as
verification binaries for eLISA, one of them being a detached double
white dwarf that shows orbital decay consistent with being only driven
by GR. The expected population is uncertain, in particular for the
interacting AM CVn systems. Once eLISA will fly this population is
measured in great detail. All of the several tens of neutron star and
black hole binaries with short orbital periods will detected by
eLISA. This rich treasure trove of data will help to unravel several
open issues in astrophysics, ranging from the progenitors of type Ia
supernovae to the structure of the Galaxy. However, before eLISA will
fly several new facilities will come on line that will contribute to
the knowledge of the Galactic binary population, so joint efforts combining
electro-magnetic as well as gravitational wave data need to be developed.

\acknowledgements It is a pleasure to thank my colleagues, in
particular Sweta Shah, Marc van der Sluys, Paul Groot,Tom Marsh, Tyson
Littenberg, Shane Larson, Neil Cornish, Michele Vallisneri, Samaya
Nissanke and Tom Prince for pleasant collaborations and interesting
discussions. This research is sponsored by the Dutch science
foundation NWO and FOM.

\bibliography{binaries}
\bibliographystyle{asp2010}

\end{document}